\documentclass[prd,twocolumn,preprintnumbers,superscriptaddress]{revtex4-2}
\usepackage{amsmath}
\usepackage{color}
\usepackage{ulem}
\usepackage{graphicx}
\usepackage{hyperref}
\usepackage{blindtext}
\hypersetup{colorlinks, citecolor=blue, linkcolor=blue, urlcolor=magenta}
\usepackage{soul} 
\usepackage[caption=false]{subfig}
\usepackage{amssymb} 
\usepackage[american]{babel}
\usepackage{microtype}
 \usepackage[table,xcdraw]{xcolor}
\definecolor{darkblue}{rgb}{0.0,0.0,0.6}

\usepackage{geometry}
\usepackage{fullpage}
\begin{document}

\title{Big Bang Nucleosynthesis constraints on the Energy-Momentum Squared Gravity: The $\mathbb{T}^{2}$ model}
\author{Dukjae~Jang}
\email{djang@buaa.edu.cn}
\affiliation{
School of Physics, Peng Huanwu Collaborative Center for Research and Education, and International Research Center for Big-Bang Cosmology and Element Genesis, Beihang University, Beijing 100191, China
}

\author{Mayukh R. Gangopadhyay}
\email{mayukh$\_$ccsp@sgtuniversity.org}
\affiliation{Centre For Cosmology and Science Popularization (CCSP), SGT University, Gurugram, Delhi- NCR, Haryana- 122505, India}
\author{Myung-Ki~Cheoun}
\email{cheoun@ssu.ac.kr}
\affiliation{Department of Physics and OMEG Institute, Soongsil University, Seoul 156-743, Republic of Korea}
\author{Toshitaka Kajino}
\email{kajino@buaa.edu.cn}
\affiliation{School of Physics, Peng Huanwu Collaborative Center for Research and Education, and International Research Center for Big-Bang Cosmology and Element Genesis, Beihang University, Beijing 100191, China \\ The University of Tokyo, 7-3-1 Hongo, Bunkyo-ku, Tokyo 113-0033, Japan,\\ National Astronomical Observatory of Japan 2-21-1 Osawa, Mitaka, Tokyo 181-8588, Japan}
\author{M. Sami}
\email{sami$_$ccsp@sgtuniversity.org, }
\affiliation{Centre For Cosmology and Science Popularization (CCSP), SGT University, Gurugram, Delhi- NCR, Haryana- 122505, India\\
Center for Theoretical Physics, Eurasian National University, Astana 010008, Kazakhstan\\
Chinese Academy of Sciences,52 Sanlihe Rd, Xicheng District, Beijing.
}
\begin{abstract}
Scale-independent energy-momentum squared gravity (EMSG) allows different gravitational
couplings for different types of sources and has been proven to have interesting implications in
cosmology. In this paper, the Big Bang Nucleosynthesis (BBN) formalism and the latest observational
constraints on nuclear abundances are being used to put bounds on this class of modified gravity models.
Using the tight constraint from BBN on the correction term in the Friedmann equation in EMSG
scenario, we report the allowed deviation from the standard cosmic expansion rate.
\end{abstract}
\maketitle

\section{Introduction}\label{intro}

The idea of modification of Albert Einstein's General Relativity (GR) \cite{Einstein1, Einstein2} dates back to the first few months after the seminal paper published by Einstein. The proposals were made to extend the GR and incorporate it into a larger, more unified theory. A few examples are Eddington’s theory of connections, Weyl’s scale-independent theory, and the higher dimensional theories of Kaluza and Klein. However, even after 108 years, the field equations proposed by Einstein remain the best description of how space-time behaves on macroscopic scales.
Einstein's equations govern everything that happens in our universe, from its expansion, structure formation, and black holes to the propagation of gravitational waves. Nevertheless, efforts to extend or modify GR never stopped, simply to understand the dynamics of dark energy (DE) and dark matter (DM). To make a comprehensive list of such models, the reader can go through the following to understand the motives and development of such theories: \cite{BD, mimetic, teves, mog,skordis, faraoni, mashhoon}. The recent curve ball thrown to us by the universe is dubbed the $H_0$ tension, and the $S_8$ tension also points towards the requirement of some modification in the GR or some extension. Although in this paper we are not focusing on the solution of the $S_8 $ tension, a modification of the cosmic expansion history can have interesting implications that need further studies in this regard.

There exists a specific class of modified theories that permit the presence of scalars constructed from the energy-momentum tensor $T_{\mu\nu}$ in the action. One can see it in $f(R,T)$ gravity, where the action involves the scalar $T=g^{\mu\nu}T_{\mu\nu}$, which is the trace of $T_{\mu\nu}$~\cite{harko}. The $f(R,\mathbb{T}^2)$ model has $\mathbb{T}^2\equiv T^{\mu\nu}T_{\mu\nu}$ in the action~\cite{Katirci,roshan2016energy,akarsu2017,board2017}. This model inspired by phenomenological considerations is coined as Energy-Momentum-Squared-Gravity (EMSG). A similar term is induced on the RS brane as a high-energy correction to the Einstein equations~\cite{Randall:1999ee, Randall:1999vf, Bennai:2006th, Gangopadhyay:2016qqa, Calcagni:2004bh}. 

It should be noted that most of the modifications of gravity, in particular scale modifications, involve extra degrees of freedom, which need to be screened out locally by mechanisms such as chameleons or Vainshtein. Theories that involve chameleon screening have great potential for late-time cosmology. However, proper screening consistent with local gravity constraints leaves no scope for late-time acceleration caused by large-scale modifications in this scenario.
One of the interesting features of EMSG is that, unlike most modified theories of gravity, it does not involve extra degrees of freedom.

The implications of EMSG for late-time acceleration have been studied in Refs.\,\cite{akarsu2017,board2017}. The model has also been explored within various cosmological frameworks \cite{nazari, bahamonde, akarsu11, akarsu12, nazari111,   Akarsu:2023agp, nari, kazemi, nazari2, nazari3, akarsu3, faraji, Ranjit:2020syg}. Chaotic inflation \cite{linde1983chaotic,linde1982new} has recently been examined in the framework of EMSG \cite{Mansoori11}, and then production of Primordial Black Hole (PBHs) and Primordial Gravitational Wave (PGW) is currently being studied in \cite{HosseiniMansoori:2023mqh}. It has been reported that a model like chaotic inflation (excluded in standard cosmology by observation) in the larger umbrella theory of EMSG falls well within the allowed limits of Planck'18 \cite{ade2021improved}.

Big Bang Nucleosynthesis (BBN) provides another critical testbed for evaluating the EMSG model. A modification of the cosmic expansion rate during the BBN epoch, caused by modified gravity models, can affect the primordial abundances of light elements. Accordingly, constraints on various modified gravity models in BBN have been investigated \cite{bbn1, bbn2, bbn3, bbn4, bbn5, bbn6, bbn7}. Thus, subjecting the EMSG model to the rigorous test of BBN is essential for providing robust insights into primordial cosmology and its phenomenology within this framework of modified gravity. In this paper, we present the effects of the EMSG model on BBN and derive constraints on the model based on BBN observations.

The paper is organized as follows: Section~\ref{bckgrnd} provides a short synopsis of the EMSG model and finally introduces the modified Friedmann equation, which plays the main role in this analysis. The stability criterion is also discussed. In Section~\ref{bbn}, we constrain the EMSG model using the latest observations of BBN. Interestingly, we have shown that BBN itself demands the negative value of the model parameter $\alpha$, which is required for a stable solution in this framework. Finally, we conclude with our findings and future directions in this context in the last Section~\ref{conc}. 

\section{Background Equations and the EMSG Model}\label{bckgrnd}

The action of the general Energy Momentum Powered Gravity (EMPG) model  is given by~\cite{ akarsu2017,board2017}:
\begin{eqnarray}\label{action}
S = \frac{\kappa}{2}\int d^{4}x &\sqrt{-g}&\left[ M_{\text{\rm p}}^4 \left( R - 2 \Lambda \right) \right. \\ \nonumber
&&- \left. \alpha\, M_{\rm p}^{4(2 \beta - 1}) (\mathbb{T}^{2})^{\beta} \right]  + \int d^4 x \sqrt{-g} \mathcal{L}_{\rm m},
\end{eqnarray}

 where $\kappa = 8 \pi G$, $M_{\rm p}$ is the reduced Planck mass, $R$ is the Ricci scalar associated with the spacetime metric $g_{\mu \nu}$, $\Lambda$ is the cosmological constant, and $\mathcal{L}_{\rm m}$ is the Lagrangian density corresponding to the matter source described by the energy-momentum tensor $T_{\mu \nu}$. Here, and in all that follows, we use units in which $\hbar = c= k_B = 1$. Then, taking $\beta = 1$ in Eq.\,(\ref{action}), one gets the EMSG action as follows
 \cite{roshan2016energy}:
\begin{equation}
S = \frac{1}{2\kappa}\int d^{4}x \sqrt{-g}\left[ R  - 2 \Lambda - \alpha\, (\mathbb{T}^{2}) \right]+ \int d^4x \sqrt{-g} \mathcal{L}_{\rm m},
\end{equation}
 where $\mathbb{T}^2\equiv T_{\mu \nu}T^{\mu \nu}$ is a scalar and $\alpha$ is a dimensionful constant that determines the coupling strength of the EMSG modification, since we are interested in the early universe in this paper, we neglect the cosmological constant, $\Lambda$.

The details of the development of the background theory of this model can be found in \cite{roshan2016energy, akarsu2017,board2017}. For the EMSG model, the effective Einstein field equation can be written as:
\begin{equation}
    G_{\mu\nu} + \Lambda g_{\mu\nu}= \kappa T_{\mu \nu}^{eff}~,
    \label{fldeqn}
\end{equation}

where $G_{\mu\nu}$ is the Einstein tensor and the effective energy-momentum tensor is given by:
\begin{equation}
    T_{\mu \nu}^{eff}= T_{\mu \nu} + \frac{2 \alpha}{\kappa}  \left ( \Psi_{\mu \nu} + T^{\sigma}_{\mu}T_{\nu\sigma} - \frac{1}{4}g_{\mu \nu} T_{\alpha \beta}T^{\alpha \beta} \right )~,
\end{equation}
where
\begin{equation}
    \Psi_{\mu \nu}= T^{\alpha\beta}\frac{\delta T_{\alpha\beta}}{\delta g^{\mu\nu}}~.
\end{equation}

Let us assume flat Friedmann-Lema\^itre-Robertson-Walker geometry:
\begin{equation}
    ds^2 = -dt^2+ a(t)^2 (dx_i^2)~, 
\end{equation}
where $i$ runs from $1$ to $3$, and $a(t)$ is the usual scale factor. Considering the perfect fluid with ideal energy-momentum tensor $T_{\mu\nu}= (\rho + p)u_{\mu} u_{\nu} + p g_{\mu\nu}$, where $\rho$ is the energy density, $p$ is the pressure, and $u_{\mu}$ is the four-velocity of fluid. Then, using Eq.\,(\ref{fldeqn}), one can use the Eq.\,(\ref{fldeqn}) to find the modified Friedmann equation as follows \cite{roshan2016energy}:
\begin{equation}
H^2=\frac{\kappa}{3} \rho -\alpha\left(\frac{1}{2}p^2+\frac{4}{3}\rho p+\frac{1}{6}\rho^2\right)~,
\label{eq_H}
\end{equation}
here $H= \dot{a}/a$ is the usual Hubble parameter. The modification term with the constant $\alpha$ in front remains non-negligible deep into the radiation domination and thus can impact BBN. Thus, the BBN constraints become very important to look into while making any claim in this domain. One can study the dynamics associated by keeping in mind that the equation of state follows that of radiation for this study. $\alpha$ is a dimensionful quantity in our analysis. Though in our paper we did not consider any particular model of inflation or its consequences, in general, to keep the gradient instability condition in mind, we have worked in the negative value of $\alpha$.
Interestingly, when BBN constraints are imposed, $\alpha$ having a negative value is also the requirement to match the current bounds. That is discussed in detail in the next section.

\section{Constraints from BBN}\label{bbn}
To evaluate the energy density and pressure in Eq.\,(\ref{eq_H}) during the BBN epoch, we consider ordinary species such as photon ($\gamma$), neutrinos ($\nu$ and $\bar{\nu}$), electron ($e^-$), positron ($e^+$), and baryon ($b$). As a result, the total energy density and pressure are, respectively, written as: 
\begin{eqnarray}
    \rho_{\rm total} &=& \rho_\gamma + \rho_{\nu + \bar{\nu}} + \rho_{e^+} + \rho_{e^-} + \rho_b, \\ [12pt]
    p_{\rm total} &=& p_\gamma + p_{\nu +\bar{\nu}}  + p_{e^+} + p_{e^-} + p_b,
\label{eq_rho_p}
\end{eqnarray}
which are incorporated in the BBN calculation code \cite{Kawano:1992ua, Smith:1992yy}. For photons and neutrinos, each energy density is given as \cite{Wagoner:1967}:
\begin{eqnarray}
    \rho_\gamma &=&  \frac{\pi^2}{15} T^4,   \\ [12pt]
    \rho_{\nu+\bar{\nu}}  &=&  \frac{7}{8} \left( \frac{\pi^2} {15} \right) N_\nu T_\nu^4  , 
\end{eqnarray}
where $k_B$ is the Boltzmann constant, $N_\nu$ is the number of neutrino species, taken as 3, and $T_\nu$ is the neutrino temperature. For these massless species, pressure is given by $p_\gamma = \rho_\gamma/3$ and $p_{\nu(\bar{\nu})} = \rho_{\nu(\bar{\nu})}/3$, respectively.

For electrons and positrons, we adopt the following energy density and pressure, respectively \cite{Fowler:1964zz}:
\begin{eqnarray}
    \rho_{e^\pm} &=& \frac{1}{\pi^2} \int_{m_e}^\infty \frac{E^2 (E^2 - m_e^2)^{1/2}}{\exp[(E - \mu_{e^\pm}}/T]+1 dE, \\[12pt]
    p_{e^\pm}    &=& \frac{1}{3\pi^2} \int_{m_e}^\infty \frac{(E^2-m_e^2)^{3/2}}{\exp\left[ (E - \mu_{e^\pm})/T\right] + 1} dE,
\end{eqnarray}
where $m_e$ is the electron mass and $\mu_{e^\pm}$ is the chemical potential of $e^{\pm}$. Under the relativistic condition where $T \gg m_e$, $p_{e^\pm} = \rho_{e^\pm}/3$. On the other hand, if $T \lesssim m_e$, the equation of state would differ from the value. During the BBN epoch, the equation of state for electrons and positrons transitions from a relativistic to a non-relativistic regime. This change affects the overall behavior of the total equation of state, as described below.

For baryons, since relativistic particles predominantly contribute to the energy density and pressure due to the enough high temperature in the BBN epoch, the contribution is negligible to the total energy density and pressure. This is further supported by the baryon-to-photon ratio: $\eta \sim 10^{-10}$.

\begin{figure}
\includegraphics[width=8.5cm]{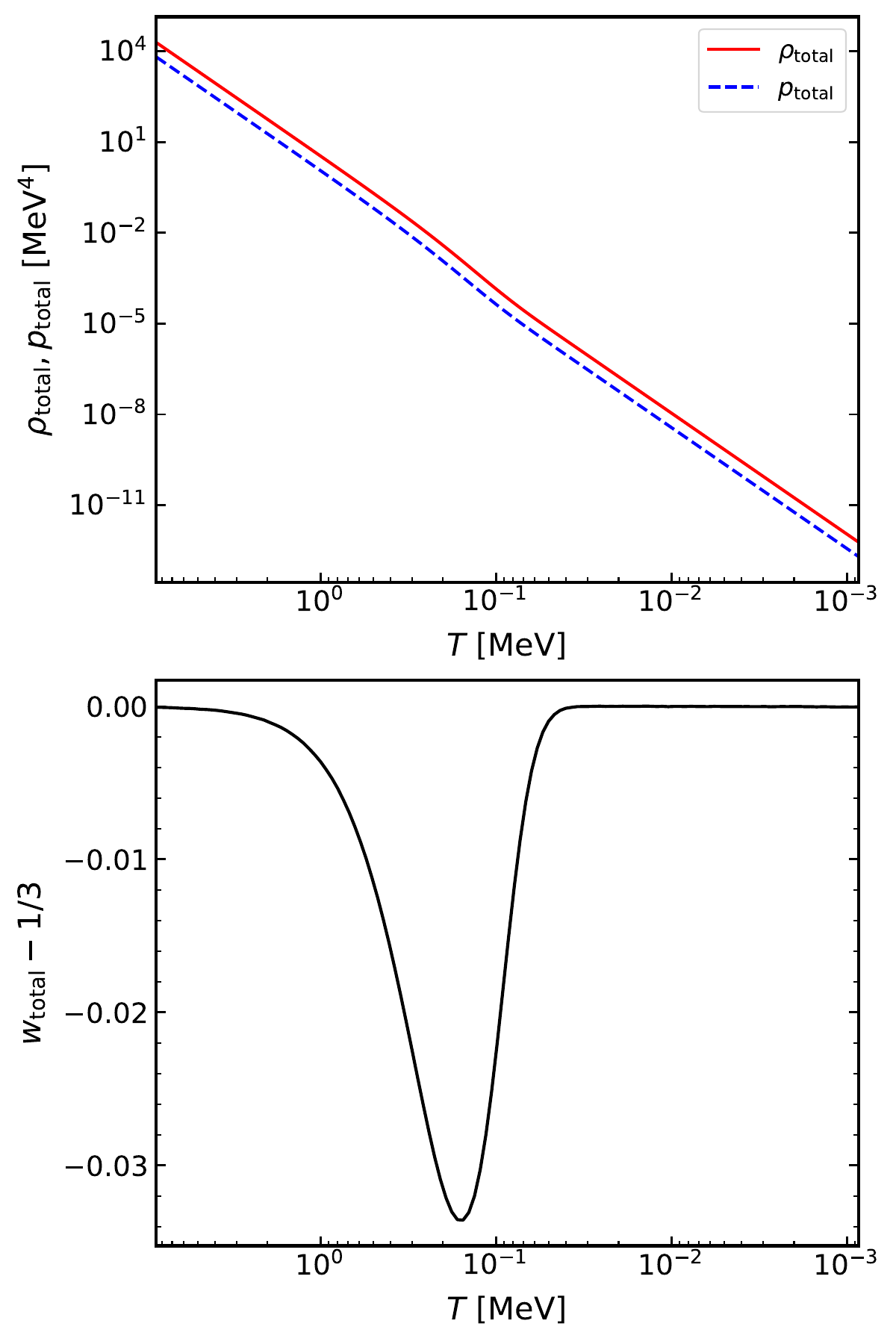}
\caption{The evolution of total energy density, total pressure, and equation of state during the BBN epoch. In the upper panel, red solid and blue dashed lines indicate the total energy density and pressure, respectively. The lower panel shows the deviation of $w_{\rm total}$ ($=p_{\rm total}/\rho_{\rm total}$) from 1/3, which represents the equation of state for radiation-like particles.}
\label{fig_eos}
\end{figure}

Fig.\,\ref{fig_eos} shows $\rho_{\rm total}$, $p_{\rm total}$, and $w_{\rm total}$ during the BBN epoch, where $w_{\rm total} \equiv p_{\rm total}/\rho_{\rm toatal}$. In Fig.\,\ref{fig_eos}, the total energy density and pressure are proportional to $T^4$ because the energy density and pressure are dominated by relativistic species. The notable point is the change in the equation of state around $T \sim 0.1\,{\rm MeV}$, attributed from the electron-positron pairs. In the early stage, for the radiation-like particles, $w_{\rm total}$ remains constant at 1/3. In this region, positrons are also relativistic due to the condition of $T \gg m_e$. However, as the temperature decreases, the equation of state for $e^\pm$ deviates from a radiation-like behavior, causing the $w_{\rm total}$ to deviate from 1/3. Subsequently, as $T \ll m_e$, electron-positron pairs become matter-like, and those contributions can be negligible in the total energy density and pressure. Consequently, the $w_{\rm total}$ returns to 1/3. 

Such a change in the equation of state shown in Fig.\,\ref{fig_eos} is one of the characteristics in the BBN epoch, stemming from the transition of electron-positron pairs from relativistic to non-relativistic species. This transition induces a temperature difference between photons and decoupled neutrinos, affecting their respective energy densities and pressures. Furthermore, since the change in the equation of state affects the continuity equation, the time-temperature relation differs from the one derived under the constant equation of state $p/\rho=1/3$. Hence, we incorporated the modified cosmic expansion rate, along with the total energy density and pressure, into our BBN calculation code to simultaneously compute the cosmic expansion rate and BBN abundances.

\begin{figure} [t]
\centering
\includegraphics[width=8.5 cm]{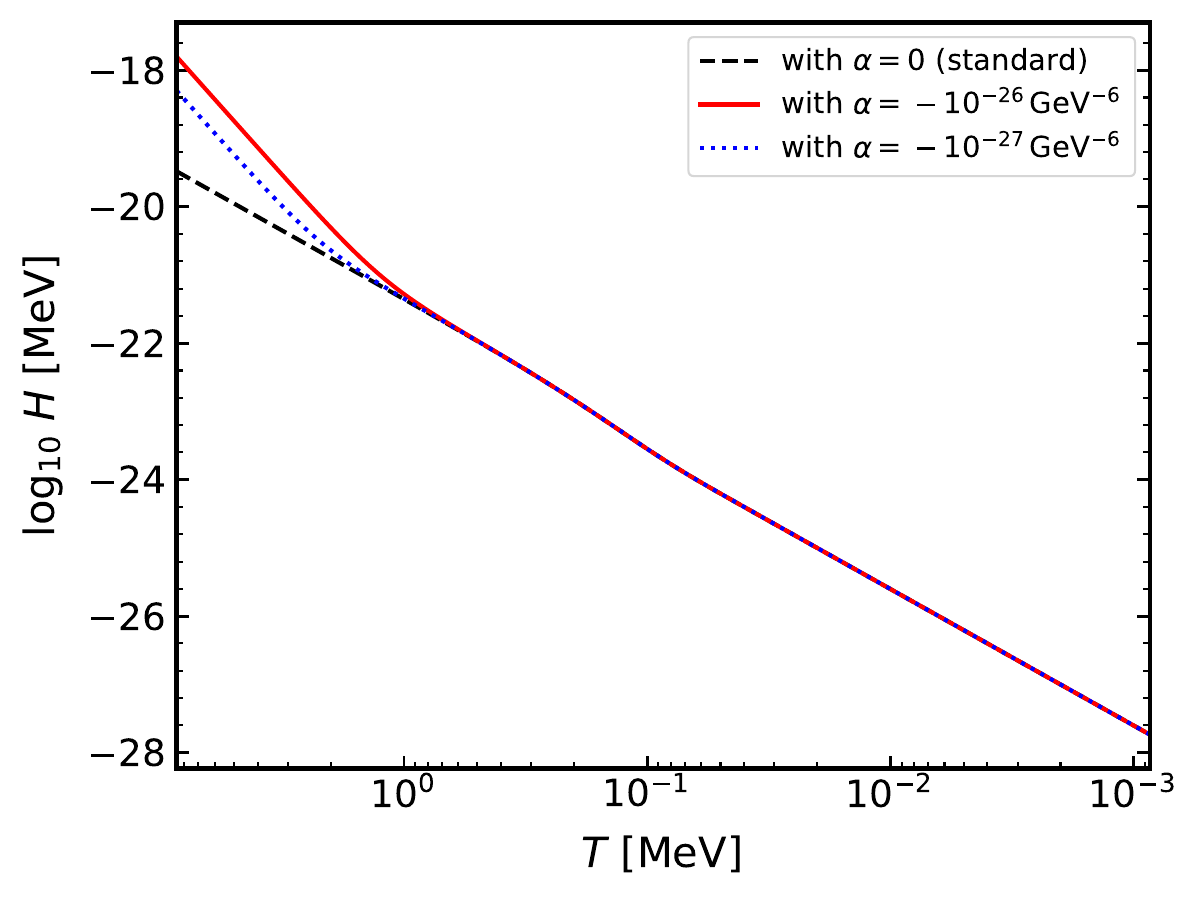}
\caption{The cosmic expansion rate in Eq.\,(\ref{eq_H}) during the BBN epoch. The red-solid, blue-dotted and black-dashed lines indicate the $H$ for $\alpha=-10^{-26}\,{\rm GeV^{-6}}$}, $\alpha=-10^{-27}\,{\rm GeV^{-6}}$, and $\alpha=0$ (standard), respectively.
\label{fig_H}
\end{figure}

By substituting $\rho_{\rm total}$ and $p_{\rm total}$ into $\rho$ and $p$ in Eq.\,(\ref{eq_H}) respectively, we evaluate the modified cosmic expansion rate, $H$, in Eq.\,(\ref{eq_H}). Fig.\,\ref{fig_H} illustrates the $H$ for $\alpha = -10^{-26}\,{\rm GeV^{-6}}$ and $\alpha = -10^{-27}\,{\rm GeV^{-6}}$, comparing it with the standard formula for $\alpha=0$. A notable deviation in the $H$ is observed in the high-temperature region. For $T \ge 1\,{\rm MeV}$, the correction term in Eq.\,(\ref{eq_H}) involving $\alpha$ can be significant compared to the first term. This is because the first term in Eq.\,(\ref{eq_H}) is proportional to $\rho$ ($\propto T^4$), while the correction term is proportional to squared energy-momentum terms ($\propto T^8$). However, due to their proportionality, both squared energy and pressure terms decrease more rapidly with decreasing temperature compared to the first term. This rapid decay of the correction term leads to a smaller correction to the $H$. Therefore, for specific $\alpha$, the modified $H$ in the EMSG model impacts the initial stage of BBN at $T \sim 1\,{\rm MeV}$, and subsequently converges with the standard cosmic expansion rate in the $T \lesssim 0.1\,{\rm MeV}$ region.

Taking into account the modified $H$, we perform the BBN calculation. We employ the BBN calculation code \cite{Kawano:1992ua, Smith:1992yy}  with updated reaction rates from the JINA REACLIB database \cite{Cyburt:2010}. As input parameters, we adopt the central value of the neutron mean lifetime provided by the Particle Data Group, $\tau_n = 878.6 \pm 0.6\, s$ \cite{ParticleDataGroup:2022pth}, and the lower limit of the baryon-to-photon ratio, $\eta = (6.104 \pm 0.058) \times 10^{-10}$, which corresponds to the baryon density based on the $\Lambda$CDM model (TT, TE, EE+lowE) from Planck observations of the cosmic microwave background, $\Omega_bh^2 = 0.02230 \pm 0.0021$ \cite{Planck:2018vyg}.

\begin{figure} [t]
\centering
\includegraphics[width=8.5 cm]{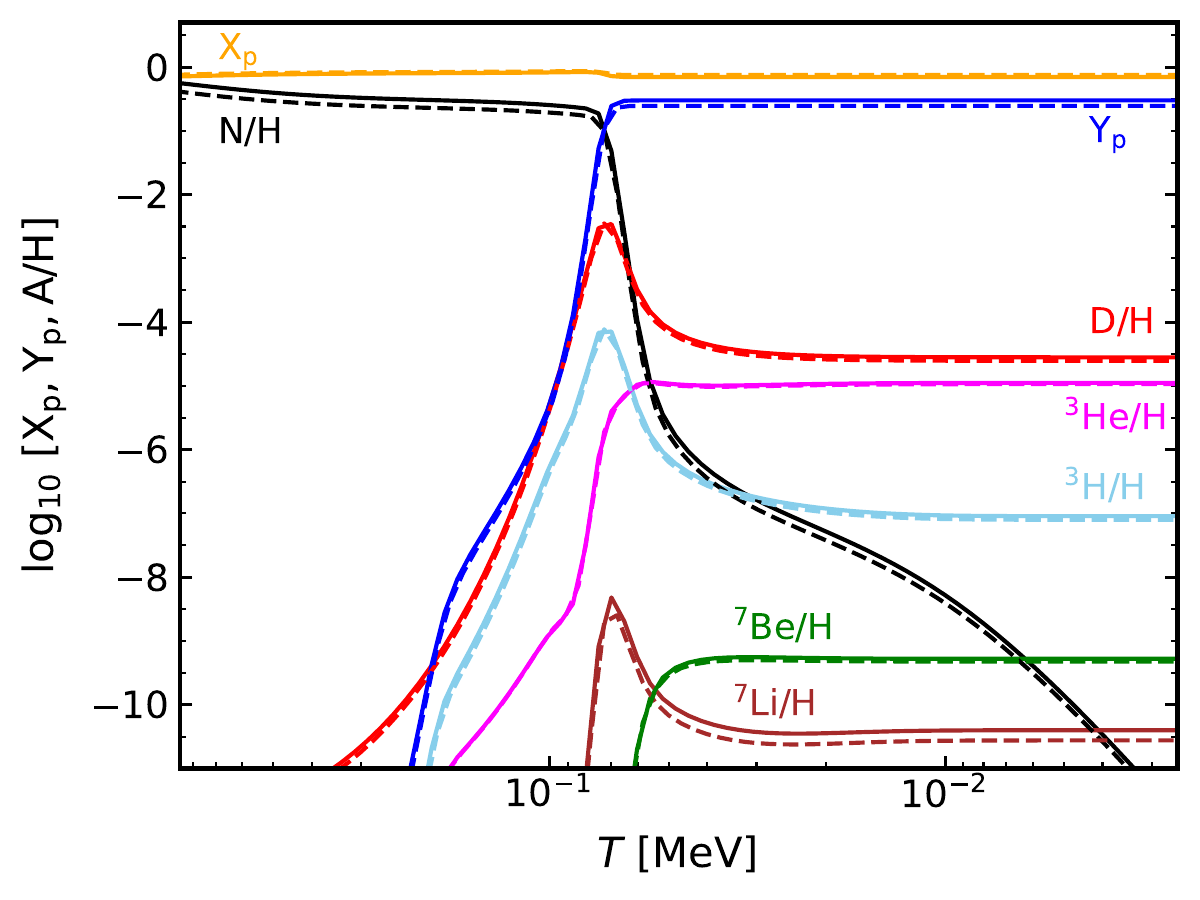}
\caption{Evolution of primordial abundances as a function of temperature. Solid and dashed lines indicate the results with $\alpha=- 10^{-25}\,{\rm GeV^{-6}}$} and $\alpha=0$ (standard), respectively. ${\rm X_p}$ denotes the mass fraction of proton, ${\rm Y_p}$ the mass fraction of $^4{\rm He}$, and ${\rm A/H}$ the abundances of element A labeled in the figure. ($N$ stands for the neutron.)
\label{fig_ab}
\end{figure}

Fig.\,\ref{fig_ab} shows the evolution of primordial abundances with $\alpha=-10^{-25} {\rm GeV^{-6}}$ and standard case of $\alpha=0$.
As shown in Fig.\,\ref{fig_H}, the negative $\alpha$ values result in an increased $H$ at the early BBN epoch, leading chemical equilibrium between neutrons and protons to freeze out earlier. As a result, the neutron abundance in the EMSG model is higher than that obtained by the standard BBN calculation, as shown in Fig.\,\ref{fig_ab}.  This increased neutron abundance in the EMSG model enhances the deuterium (D) abundance through the $^1{\rm H}(n,\gamma)^2{\rm H}$ reaction. Then, a larger D also increases abundances of $^3{\rm H}$ and $^3{\rm He}$ through ${\rm D}(d,p)^3{\rm H}$ and ${\rm D}(d,n)^3{\rm He}$ reactions, respectively, which increase $^4{\rm He}$ abundance by enhancing $^3{\rm H}(d,n)^4{\rm He}$ and $^3{\rm He}(d,p)^4{\rm He}$ reactions. Consequently, abundances of D, $^3{\rm He}$, and $^4{\rm He}$ increase as $\alpha$ decreases. This trend aligns with findings from other studies on the effects of a modified expansion rate on primordial abundances \cite{Kusakabe:2015ida, Jang:2016rpi, Sasankan:2017eqr, Mathews:2017gbj, Mathews:2018qei}.

\begin{figure} [htb!]
\centering
\includegraphics[width=8.5cm]{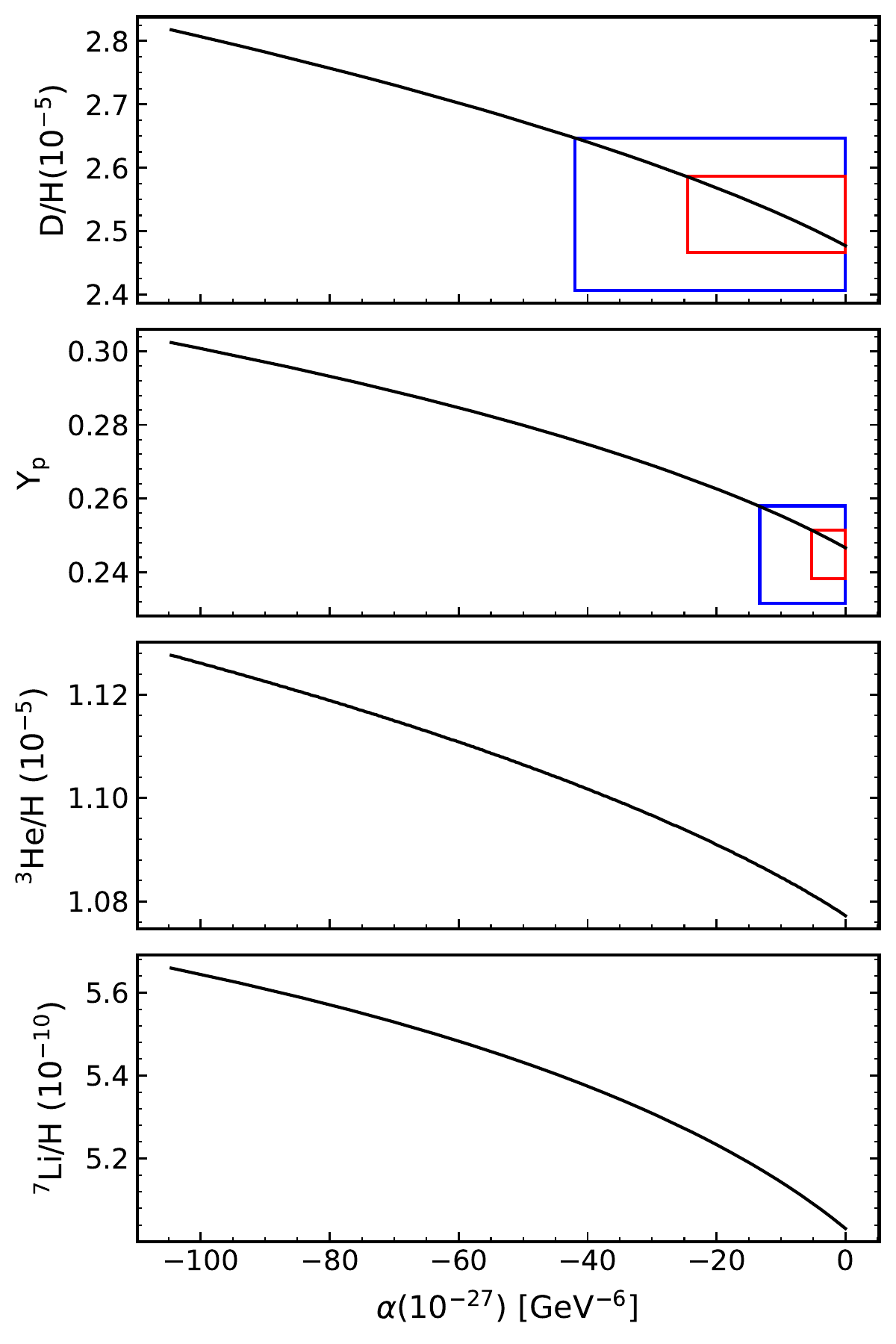}
\caption{Final abundances of of D, $^4{\rm He}$, $^3{\rm He}$, and $^7{\rm Li}$ as a function of $\alpha$. In the first and second panels, the red and blue boxes indicate constrained regions by the observational data within $2\sigma$ and $4 \sigma$ range, respectively. We adopted the observational data of D abundance from the metal-poor Lyman-$\alpha$ absorption, ${\rm D/H}= 2.527 \pm 0.030$ \cite{Cooke:2017cwo}, and mass fraction of ${\rm ^4He}$ from metal-poor extra-galactic H II regions, ${\rm Y}_p= 0.2448 \pm 0.0033$ \cite{Aver:2022}. For ${\rm ^3He}$ in the third panel, the abundance is consistent with the observational upper limit of $^3{\rm He/H} = 1.1 \pm 0.2$ \cite{Bania:2002yj}. In the fourth panel, the abundance of ${\rm ^7Li}$ for the given $\alpha$ is higher than the observational data of $^7{\rm Li/H}=1.58 \pm 0.31$ \cite{Sbordone:2010}. 
}
\label{fig_H2}
\end{figure}

Fig.\,\ref{fig_H2} depicts the final abundances of D, $^3{\rm He}$, $^4{\rm He}$, and $^7{\rm Li}$ as a function of $\alpha$. As mentioned above, a decreased $\alpha$ increases neutron-to-proton ratio, which leads abundances of D/H, $^3{\rm He}$, and $^4{\rm He}$ to increase. For D abundance, compared to the observational data from the metal-poor Lyman-$\alpha$ absorption \cite{Cooke:2017cwo}, we find the constraint region of $\alpha$ to be $-2.45 \times 10^{-26}\,{\rm GeV^{-26}}$ ($2\sigma$) and $-4.20 \times 10^{-26}\,{\rm GeV^{-6}}$ ($4\sigma$). For a mass fraction of ${\rm ^4He}$ from metal-poor extra-galactic H II regions \cite{Aver:2022}, we find that the constrained region is narrower, with values of $-5.25 \times 10^{-27}\,{\rm GeV^{-6}}$ ($2\sigma$) and $-13.3 \times 10^{-27}\,{\rm GeV^{-6}}$ ($4\sigma$). For the abundance of ${\rm ^3 He}$, it is relatively insensitive, so that all regions of $\alpha$ are allowed by the upper limit of observational data of $^3{\rm He/H} = 1.1 \pm 0.2$ \cite{Bania:2002yj}.  

We also discuss $^7{\rm Li}$ abundance in the EMSG model. The increased $^{3}{\rm H}$ and $^{3}{\rm He}$ due to the negative $\alpha$ enhance $^3{\rm H}(\alpha, \gamma)^7{\rm Li}$ and $^{3}{\rm He}(\alpha, \gamma)^7{\rm Be}$ reactions, which lead to increase in $^7{\rm Li}$ and $^7{\rm Be}$. In particular, the final abundance of $^7{\rm Be}$ mainly contributes to the final $^7{\rm Li}$ abundance by radioactive electron capture. Consequently, the abundance of $^7{\rm Li}$ increases as $\alpha$ decreases, which is shown in Fig.\,\ref{fig_ab}. This implies that changes in the $\alpha$ deepen the over-prediction of primordial lithium abundance in the standard BBN (SBBN) model.

\section{Conclusion}\label{conc}
 We have explored the impact of the EMSG model on BBN and its implications. For negative values of $\alpha$, the correction term in the EMSG model enhances the cosmic expansion rate, depending on the squared energy density and pressure. Constraints on the parameter $\alpha$ were obtained earlier using cosmic microwave background (CMB) and baryonic acoustic oscillation (BAO) data \cite{akarsu11}; here we have focused on the implications due to the demands of BBN.
Given the radiation domination due to relativistic species during the BBN epoch, the correction term rapidly decays over cosmic time, proportional to $T^8$. Consequently, the EMSG correction specifically affects the initial stage of the BBN epoch, which leads to an increase in the primordial abundances of D and $^4$He for a larger negative value of $\alpha$.

Using the BBN observations, we have shown that the lower limits of $\alpha$ are constrained to $-5.25 \times 10^{-27}\,{\rm GeV^{-6}}$ and $-13.3 \times 10^{-27}\,{\rm GeV^{-6}}$ within $2 \sigma$ and $4\sigma$ ranges, respectively.

One can find that the BBN bounds are much more stringent than the combined CMB+ BAO bounds, as reported in \cite{akarsu11}. 
There are several paths to explore after this, one being keeping the inflationary physics in mind, and how the BBN will have more stringent effects in this scenario. For instance, in the case of braneworld dynamics, the demands of inflationary observations put quite strict bounds on the brane tension as reported in \cite{Bhattacharya:2019ryo}. Similarly, it would be valuable to investigate the effect on EMSG when inflationary observables are taken into account.

One important aspect that came of the analysis is that BBN bounds prefer the negative value of $\alpha$ which is also the theoretical demand to avoid ghost and gradient instability in this model with $\beta= 1$. 
It is interesting to note that the Friedmann equation, in this case, has a similarity with what one expects in the case of loop quantum gravity or the braneworld cosmology. In general, one can say that BBN will have the most stringent bound with respect to the other tests of these theories.

Finally, we emphasize that any modification to Einstein's General Relativity, which can affect the BBN, has to satisfy the stringent constraints imposed by BBN observations. Moreover, such modifications should be further developed to ensure their consistency with cosmological observations at later times.

\section{Acknowledgement}
Work of MRG is supported by the Department of Science and Technology(DST), Government of India under the Grant Agreement number IF18-PH-228 (INSPIRE Faculty Award) and by the Science and Engineering Research Board (SERB), DST, Government of India under the Grant Agreement number CRG/2022/004120 (Core Research Grant). 

MS is supported by the Science
and Engineering Research Board (SERB), DST, Government of India under the Grant
Agreement number CRG/2022/004120 (Core Research Grant). MS is also partially supported by the Ministry of Education and Science of the Republic of Kazakhstan, Grant
No. AP14870191 and CAS President’s International Fellowship Initiative(PIFI).

TK and DJ are partly supported by the National Natural Science Foundation of China (No. 12335009), and the National Key R$\&$D Program of China (2022YFA1602401). TK is also supported in part by Grants-in-Aid for Scientific Research of Japan Society for the Promotion of Science (20K03958).

MKC's work was supported by the Basic Science Research Program of the National Research Foundation of Korea (NRF) under Grants No. 2021R1A6A1A03043957 and No. 2020R1A2C3006177.

MRG also wants to thank the organizers of OECE,2023, Beihang University for the hospitality received when the project was initiated. 

We would also like to thank the anonymous referee for their invaluable comments and suggestions, especially regarding one mistake in the previous version of the manuscript which is now removed. 

\end{document}